\documentclass[10pt]{article}

\usepackage{comment}
\usepackage[utf8]{inputenc}
\usepackage{amsmath}
\usepackage{amsfonts}
\usepackage{amssymb}
\usepackage{mathtools}
\usepackage{graphics}

\begin{document}
	
\title{Tourist route optimization in the context of Covid-19 pandemic}
\date{}

\author{Cristina Maria Păcurar \footnote{Faculty of Mathematics and Computer Science, Transilvania University of Bra\c sov, Bulevardul Eroilor 29, Bra\c sov; cristina.pacurar@unitbv.ro}, Ruxandra-Gabriela Albu \footnote{Faculty of Economic Sciences and Business Administration, Transilvania University of Bra\c sov, Bulevardul Eroilor 29, Bra\c sov; ruxandra.albu@unitbv.ro} and Victor-Dan Păcurar \footnote{Faculty of Silviculture and Forest Engineering, Transilvania University of Bra\c sov, Bulevardul Eroilor 29, Bra\c sov; vdpacurar@unitbv.ro}}

\maketitle

\textbf{Abstract:} The paper presents an innovative method for tourist route planning inside a destination. The necessity of reorganizing the tourist routes within a destination comes as an immediate response to the Covid-19 crisis. The implementation of the method inside tourist destinations can bring an important advantage in transforming a destination into a safer destination in times of Covid-19 and post-Covid-19. The existing trend of shortening the tourist stay length has been accelerated while the epidemic became a pandemic. Moreover, the wariness for future pandemics has brought into spotlight the issue of overcrowded attractions inside a destination at certain moments. The method presented in this paper proposes a backtracking algorithm, more precisely an adaptation of the travelling salesman problem. The method presented is aimed to facilitate the navigation inside a destination and to revive certain less-visited sightseeing spots inside a destination while facilitating conformation with the social distancing measures imposed for Covid-19 control.


\section{Introduction}

At the end of December 2019, a new strain of coronavirus emerged in Wuhan, China. The new coronavirus, named SARS-CoV-2 in February 2020, causes the Covid-19 disease, which has affected, at a steady pace, the entire world. Currently, almost every country in the world has reported cases of Covid-19. The pandemic has affected in a different proportion every industry and sector. Among them, one of the industries which is severely and most probably long-term affected is the tourism industry.

The effects of the Covid-19 pandemic on tourism can be acknowledged at many different levels. The measures taken worldwide to prevent the spreading of the disease have affected all possibilities of travel and forced many people to cancel their trips or postpone them to a later unknown date. 

The lock-downs and numerous restrictions imposed by governments have thrown many tourism-related businesses into collapse. Therefore, since international travel still comes with significant restrictions and poses great threats in many places around the world, the revival of tourism will mostly relate to national, or regional, maybe even local tourism. 

On the other hand, the Covid-19 pandemic will leave significant marks on the way people travel. The most visited spots in a destination, which are usually overly crowded, require special attention and immediate reorganization. 

Travel restrictions will continue to revolutionize the idea of tourism as we know it, both from the point of view of the providers of tourism, as well as from the tourists' point of view.

We propose a method for organizing tourist routes inside a destination in accordance with the objective of shortening the tourist stay length, which has been an increasing trend even before the Covid-19 pandemic. For the local community, the implementation of this method would bring advantages related to each of the three sustainable development dimensions, namely the reduction of the negative effects brought by overcrowding the destination, the possibility of continuing the tourism business, while the current global crisis generated by Covid-19 is ongoing and altogether maintaining the incidence rate of the new coronavirus low and ensuring the inhabitants' health. 

The method we propose is based on a backtracking algorithm to find the optimal route to take for visiting a certain number of attractions inside a destination. In order to find this optimal route (in terms of minimal length or duration), we adapted the algorithm used for the traveling salesman problem.

\section{Literature review}

According to Chang et al. (see \cite{Chang}) the future of tourism in the COVID-19 era is uncertain, the pandemics affecting not only humans' healthiness but even more the global economy and the very fabric of society. Consequently, there is a critical need for new research aimed to identify the best solutions for reviving the tourism industry, one of the economic sectors highly affected by the present sanitary crisis.
 
The unique relationship between tourism and sustainable development derives from the outstanding peculiarity of this economic activity that sells the physical and human environment as its product. In our opinion, there is one certain continuity element, as important in the future as it was before the COVID pandemics' outbreak, the need for tourist destination management based on sustainable development principles. Overcrowding of certain destinations is an old problem in tourism, changing in time the structural types of visitors and finally spoiling the attractivity of this destination (see \cite{Butler,Agarwal,Szromek,Szromek-2}).

The appropriate spatial planning and its subsequent tourist flux optimisation play a key role in preventing or reducing tourist overloading and its negative effects.

The problems of overcrowding and extremely high concentration of human traffic are increased to a peak in urban areas, and these are referring to both tourism and anti-pandemics regulations. Cities are hot spots in the present Coronavirus crisis, being associated with the highest risk of infection and, consequently, these destinations are confronting the strongest reduction of visitors and, most probably, here, the effects of the sanitary crisis, especially on people's mentality, will be long-lasting.

Within the sustainable development paradigm, tourism is one peculiar industry, dependent on the natural and anthropic potential and the cultural heritage as well. Tourism sells these assets, but at the same time, shares these and additional resources with other stakeholders, among whom local communities are of utmost importance. It is essential for tourism to be actively involved in sustainable development and to cooperate with the other industries for preserving the quality of those resources essential for tourism activity. Sustainable development is basically an intrinsic necessity for tourism, involving the need of reducing the negative economic, social or ecological effects and their mitigation can only be achieved through professional management, which attracts in the decision-making process all the factors engaged in the development of tourism (see \cite{Albu}). 

On the other hand, tourism can bring risks and opportunities for local communities as a study undertaken by UNWTO and IPSOS, in 2019, asserts (see \cite{ipsos}). The survey, which collected 12 000 answers from 15 countries, was aimed at a better understanding of the residents' perceptions towards city tourism, its impact and the management strategies. The survey shows that 47\% of respondents consider that in their cities there are numerous visitors, with 52\% indicating that tourism has an important beneficial economic impact (moderate or big). From the interviewed subjects, 46\% thought tourism "creates overcrowding", while 49\% indicated that measures should be implemented for better tourism management, including improved infrastructure and facilities (72\%), expanded offer of attractions, for both locals and visitors (71\%), and ensuring that the local community benefits from tourism (65\%).


Presently, the 2030 Agenda for Sustainable Development, adopted in September 2015 by the General UN Assemble, could be considered the document drawing the canvas of the global environmental consciousness. More specifically, the seventeen Sustainable Development Goals (SDG) are presenting the correspondent directions for human society evolution in a green fashion, harmonised with the principles of natural heritage preservation and natural resources sustainable management. As concerns urban areas development, among the  Agenda`s SDG’s, there is one especially targeting it (SDG  11- Sustainable cities and communities), that clearly states the urgent need for significantly transforming the way we build and manage our urban spaces, where it  is already living more than a half of human population, and this share is estimated to increase at two-thirds of all humanity (6.5 billion people) at the middle of this century  (see \cite{sustainable cities, goals}).

Based on the 2030 Agenda, one year later, at the following global summit on urbanization (the United Nations Conference on Housing and Sustainable Urban Development, at Quito, Ecuador, October 2016), the world leaders adopted the New Urban Agenda which sets global standards for sustainable urban development, requiring a new approach in cities development, from design to maintenance and lifestyle, thus implying effective cooperation of all the essential stakeholders (authorities and private sector, civil society and individual households).

Sustainable Urban Tourism has been the main subject of numerous studies (see \cite{Aall, Ko, Ashworth, Lerario, Martinez, Zamfir, Sunu, Insch}). The new global Sustainable Development Goals (SDGs), adopted by the World Leaders in 2015 affirm (in Goal 11) that future cities must be inclusive, safe, resilient and sustainable (see \cite{sustainable cities}). Certainly, the safety component, in the previous listing, has added new constraints related to the present pandemic crisis (social distancing etc.). One could say that the old overcrowding issues require even more enhanced attention. Tourist overloading could be induced by the differences between accommodation demand and supply, or by the spatial concentration of the lodging facilities. These problems could be tackled by infrastructure development combined with an efficient promotion of the new facilities and services, because finally the subjective approach of the potential visitors, their informed choice would make the difference. This tourist choice question is even more important as regards the tourist traffic inside the destination, highly important especially in urban areas, where the list of tourist objectives and attractions could be short, and consequently, there is a potential risk of congestion points occurrence, extremely dangerous during pandemics. Most probably in the present situation the majority of tourists is highly conscious of the importance of avoiding overcrowded places. Tourists would finally decide where, when and how they will travel and visit the objectives of interest (see \cite{earth}). 

Lately, urban tourism experienced an intense development. This increased interest for visiting different cities (not only the widely known great metropolises, popular for tourists over centuries) is related to both professional travelling (for business, congresses, conferences etc.) and also to personal tourism targeting cultural, art or leisure interests.  (see \cite{Istoc}).  Tourism management could increase the demand for urban holidays, by developing new attractive locations, events, facilities, in order to draw an increased attention of domestic and international visitors (see \cite{oecd}). All those development plans must follow a sustainable development policy that can play a global role in tourism attractiveness, contributing especially to the recognition of its urban context (see \cite{Boivin}).

Tourism spatial planning has been in the spotlight for a long time and many studies have tackled this topic (see \cite{Acheampong, Almeida, Okan, Getz, Ginting, Risteski, Marzuki}).

Route planning is an area of interest for many studies, as it is a key aspect in providing (or obtaining - if regarded from the tourist’s point of view) an improved travel experience. There have been many research papers concerning route planning. The main findings in the literature regarding methods used in route planning are presented in Table \ref{LitRev}.

\begin{table}[h!]
	\caption{Tourist route planning}
	\label{LitRev}
	\begin{tabular}{ccc}
		\hline
		\	& \textbf{Method}	& \textbf{Authors, Year}\\
		\hline
		1.		& Heuristic algorithm & G.M. Hua, \textbf{2016}, \cite{li1} \\
		& 			& W. Zheng, Z. Liao, Z. Lin, \textbf{2020}, \cite{lit2} \\
		2.		& Markov Chain Model & S. Ahmad, I. Ullah, F. Mehmood, M. Fayaz, D. Kim, \textbf{2019}, \cite{lit3}\\
		3.		& GIS based algorithm &  G. Lau, \textbf{2016}, \cite{Lau} \\
		&   & N. Gill, B. Bharath, \textbf{2013}, \cite{Gill} \\
		&   & P. Du, H. Hu, \textbf{2018}, \cite{lit4}\\
		& 			& E. Abubakar, O. Idoko, O. Ocholi, \textbf{2017}, \cite{lit5}\\
		& 			& X. Zhou, Y. Zhan, G. Feng, D. Zhang, S. Li, \textbf{2019}, \cite{lit6}\\
		4.		& Machine learning & X. Zhou, M. Su, Z. Liu, Y. Hu, B. Sun, G. Feng,  \textbf{2020}, \cite{lit7}\\
		5.		& Neural networks & S. Malik, D. Kim, D., \textbf{2019}, \cite{Malik} \\
		& 			& W. Sirirak, R. Pitakaso, \textbf{2018}, \cite{lit9} \\
		6.		& Genetic algorithms & D. Perera, C. Rathnayaka, S. Dilan,  \\
		&	& L. Siriweera, W. H. Rankothge, \textbf{2018}, \cite{lit10} \\
		& 			& X. Ma, \textbf{2016}, \cite{lit11}\\
		7.		& Ant-colony algorithm & Zhang W., \textbf{2019}, \cite{lit12}\\
		& 			& 	X. Qian, X. Zhong, \textbf{2019}, \cite{lit13}\\
		& 			&  H.C. Huang, \textbf{2013}, \cite{lit14}\\
		& 			&  Ginantra, N.L.W.S.R et al. \textbf{2019}, \cite{lit15}\\
		8.		& Multiobjective &  Y. Han, H. Guan, J. Duan, \textbf{2014}, \cite{lit16} \\
		& optimization		&  \\
		9.		& MINIMAX 			&  T. Hasuike, H. Katagiri, H. Tsubaki, H. Tsuda, \textbf{2013}, \cite{Hasuike}\\
		& optimization 	& X. Wu, H. Guan, Y. Han, J. Ma, \textbf{2017}, \cite{Wu} \\
		& 			& E. Nikolova, M. Brand, D. Karger, \textbf{2006}, \cite{Nikolova}\\
		10.		& Dijskra algorithm	& Y. Xu, S. Zhang, J. Yang, \textbf{2015}, \cite{lit20}\\
		11.		& Floyd algorithm	& R. Xu, D. Miao, L. Liu, J. Panneerselva, \textbf{2017}, \cite{lit21} \\
		&	& X. Zhou, Y. Yuan, M. Ma, H. Li, H. \textbf{2018}, \cite{Zhou}\\
		\hline
	\end{tabular}
\end{table}

Even though route planning based on looking at the routes inside a destination as a graph has been used in previous studies, the \textit{shortest path} algorithms proposed in those researches (such as improved Floyd algorithm in \cite{Zhou}) result in finding the optimal route between two nodes, more precisely, the shortest path between sightseeing spots in a destination. However, tourism in smaller destinations needs to exploit all the small attractions one might find along a certain route. Thus, our method, which involves the use of a backtracking algorithm has two main useful novelties. Firstly, using the backtracking algorithm allows the establishment of an optimal path in a destination that crosses through a certain number of attractions. Secondly, the backtracking algorithm returns all the possible ways of planning the route while finding the shortest one. This is highly advantageous in the context of Covid-19, as overcrowded spots need to be reorganized, and choosing the optimal route is not only dependent on the length of the path, but it should also look at reducing the number of people present in a place, at a certain time, in order to diminish the risk of disease-spreading.

We consider that the tourist destination management structures (DMOs), must include in the destination development strategy, for the post-pandemics period, clear solutions for the safe access of tourists to the important tourist objectives, thus also enabling the small businesses to resume their economic activities at a level at least similar to that before the pandemics (if possible, increased activity would be welcome for partly compensating the 2020-2021 losses).
 
\section{Materials and Method} 

We have chosen to test our method for the city of Brașov, Romania, one of the major cities in Transilvania and the biggest in the centre region of Romania. Brașov is one of the most important economic, cultural and sports centers in the country, being a significant tourist destination and a historic city, with numerous tourist attractions, such as: the Black Church, the First Romanian School, Saint Nicholas Church, Șchei’s Gate, Catherine’s Gate, the Citadel, a.o. As a tourist destination, Brașov has experienced in the past years an increasing trend as regards the number of tourists. 

However, the current situation has seriously affected the tourist industry in Brașov, especially in the period March-May 2020, when the tourism dropped close to zero as a result of the lock-down caused by Covid-19. 

\begin{figure}[h!]
	\centering
	\includegraphics[width=8cm]{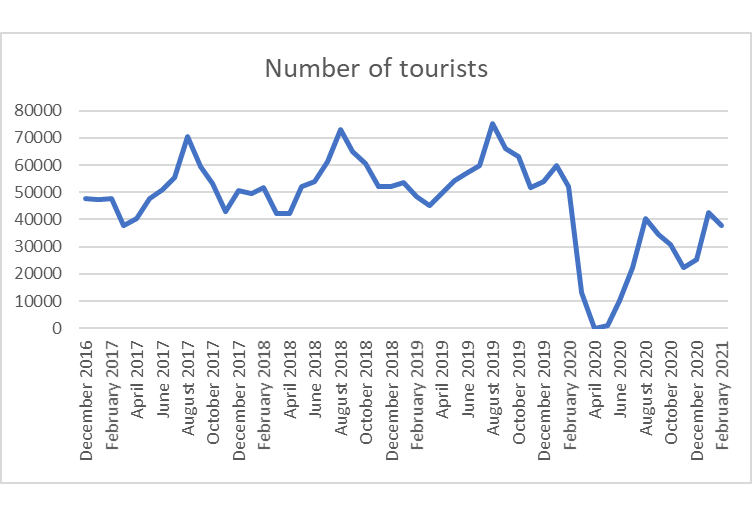}
	\caption{Effect of Covid-19 on tourism in Brasov}
	\label{Covid-ef}
\end{figure}

Like many places around the world, Brasov has been greatly affected by the Covid-19 pandemic. The decrease in the number of tourists has been significant as can be seen from Figure \ref{Covid-ef}. The graph in Figure \ref{Covid-ef} is based on data from \cite{ins}, extracted from the database TUR104H.

For a comparative study on the attractivity of Brasov before and during the pandemic see \cite{Pacurar}.

We believe that in order to face the current situation and revive tourism, new innovative methods must be identified and employed to transform the tourist offer, according to the changes in the hierarchy of the motivational elements that determine people to visit a certain destination (safety rules and other measures imposed by authorities to prevent the spread of the SARS-CoV-2 virus are the priority now).

\begin{figure}[h!]
	\begin{minipage}{0.4\textwidth}
		\includegraphics[width=\textwidth]{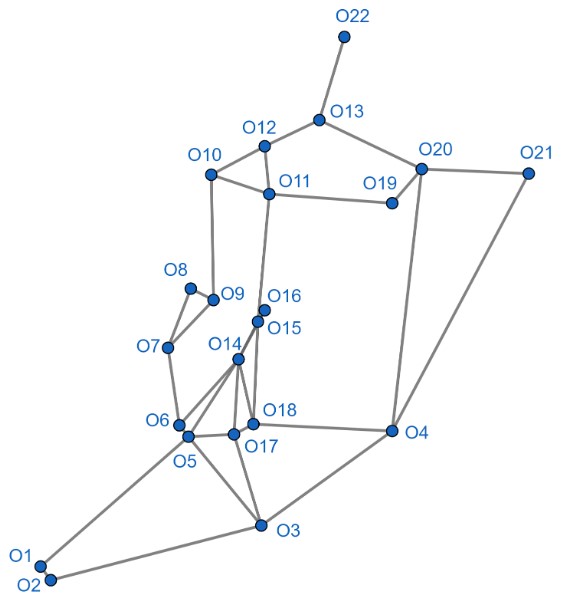}
		\caption{Graph of tourist attractions in Brașov Old City Centre}
		\label{graf}
	\end{minipage}
	\hspace{30pt}
	\begin{minipage}{0.4\textwidth}
		O1: First Romanian School \\ 
		O2: Saint Nicholas Church \\ 
		O3: Weavers' Fortress \\ 
		O4: Tâmpa Cable Way \\ 
		O5: Șchei's Gate \\ 
		O6: Catherine's Gate \\ 
		O7: Black Tower \\ 
		O8: White Tower \\
		O9: Graft Fortress \\ 
		O10: George Barițiu Library \\ 
		O11: Rectorate of \\
		\hspace{1cm} Transilvania University \\
		O12: House of Army \\ 
		O13: Annunciation Church \\ 
		O14: Black Church \\
		O15: Council Square \\ 
		O16: History Museum \\
		O17: Synagogue \\
		O18: Rope Street \\ 
		O19: Art Museum \\
		O20: Town Hall \\
		O21: Theater \\
		O22: The Citadel \\
	\end{minipage}
\end{figure}

The method employed aims to find the shortest way of passing through a number of attractions $s$ which must be chosen between 2 (which is trivial), and the maximum number of visiting objectives $n$.

The map of the attractions is considered as a not-oriented graph, which is represented in Figure \ref{graf}. The program asks for the number of vertices $n$ (in this case $n=22$ as we have considered 22 main attractions in the old city of Brasov), the number of edges $m$ (for Brașov old town, $m=35$ as can be seen in Figure \ref{graf}), the starting point and the ending node. 

The method employed is a backtracking algorithm. For further reading on this method see \cite{Cormen}. We considered the problem of finding the shortest route inside a destination that passes through a certain number of attractions as a traveling salesman problem (for further reading on this problem see \cite{Salesman}, \cite{Book}).

We describe, below, the steps of the recursive algorithm that we adapted:

\begin{itemize}
	\item[-] At each step, a new element (here, a new tourist attraction $Oi, i=\overline{1,22}$), indexed symbolically from $1$ to $n$, is introduced in the stack.
	\item[-] For every valid element at level $k$ of the stack (an element is considered valid at level $k$ of the stack if there exists an edge between it and the object existing at level $k-1$ of the stack), the compatibility with the other values of the stack is evaluated (an element is compatible with the elements existing in the stack if it is not one of the objects in levels from $1$ to $k-1$):
	\begin{itemize}
		\item[a)] if the element is compatible, it is introduced in the stack and the algorithm passes to the next step (next level of the stack);
		\item[b)] if for a certain value a solution cannot be built, the current element is dropped and another element is introduced in the stack at the current level, if there exists another element;
		\item[c)] if all elements have been tested and there is no valid solution, the lower level $k-1$ becomes the current level.
	\end{itemize}	
	\item[-] The algorithm has found a solution when the stack level is equal to the required number $n$ (for passing through all nodes from the graph of the destination), or a lower number $s$, which is initially specified, and the highest level of the stack is occupied with the \textit{stop} point fixed at the beginning. 
	
	\item[-] The algorithm has finished when all values acceptable at a certain level of the stack have been tested. When a solution is found, the distance from the starting and ending point is calculated. To obtain the shortest path, the distance obtained is firstly compared to an initial number, which is very large, and the smallest value is remembered. At every step, the distance is compared to the last value accepted (smallest number obtained until the moment of comparison).
\end{itemize}

The backtracking method presented, although it is not as efficient as the Floyd algorithm, for example, it is used when it is important to obtain all possible paths from a node $x$ to a node $y$. This is the case in the context of managing tourist routes inside a destination.

\begin{table}[h!]
	\caption{Distances between nodes in meters and minutes}
	\label{DistM}
	\centering
	\begin{tabular}{|c|c|c|c||c|c|c|c|}
		\hline
		node & node & m & min & node & node & m & min\\
		\hline
		1 & 2 & 20 & 1 & 10 & 11 & 170 & 2 \\
		1 & 5 & 650 & 8 & 10 & 12 & 180 & 2\\
		2 & 3 & 750 & 10 & 11 & 12 & 150 & 2\\
		3 & 5 & 350 & 5 & 11 & 15 & 400 & 5\\
		3 & 4 & 750 & 9 & 11 & 19 & 500 & 6\\
		3 & 17 & 350 & 5 & 12 & 13 & 230 & 3\\
		4 & 18 & 750 & 10 & 13 & 20 & 350 & 4\\
		4 & 20 & 1200 & 15 & 13 & 22 & 500 & 9\\
		4 & 21 & 1100 & 14 & 14 & 15 & 210 & 2\\
		5 & 6 & 20 & 1 & 14 & 16 & 250 & 3\\
		5 & 14 & 350 & 4 & 14 & 17 & 400 & 5\\
		5 & 17 & 110 & 2 & 14 & 18 & 350 & 5\\
		6 & 7 & 450 & 7 & 15 & 16 & 10 & 1\\
		6 & 14 & 300 & 4 & 16 & 18 & 450 & 6\\
		7 & 8 & 250 & 4 & 17 & 18 & 60 & 1\\
		7 & 9 & 260 & 4 & 19 & 20 & 300 & 4\\
		8 & 9 & 110 & 2 & 20 & 21 & 400 & 5\\
		9 & 10 & 350 & 5 & & & &\\
		\hline
	\end{tabular}
\end{table}

We implemented the algorithm in C++. For the implementation we used two files for reading the graph, with the same nodes and vertices. The first file contains the nodes and the corresponding distance between them in meters. The second file contains the nodes and the distance between them expressed in minutes. The distances in minutes are chosen according to the distances calculated by Google Maps (see \cite{Google}). The motivation for choosing two files is that for tourists the time required to move inside a destination is usually more important than the distance covered. 

The distances between nodes are synthesized in Table \ref{DistM}. This table should be read in the following order: the first four columns, line by line, until the end of lines, followed by the next four columns, which should be read line by line. The columns \textit{node} represent the index of a node. The columns \textit{m} represent the distance between the specified nodes in meters, while the columns \textit{min} represent the same distance, but calculated in minutes.

In order to empirically test the method that we propose we have conducted a qualitative research. Recommendations on the sample size in the case of qualitative research vary and some authors propose that the sample size should be in the range of 12-26 subjects (see \cite{Lubovski}). In our case, we interviewed 15 volunteers who were willing to participate in the study and test our method. The questions included in the interview guide which were addressed to the volunteers are presented in the Appendix.

\section{Results}

For obtaining the first results, we used, as input files, text files containing the values in Table \ref{DistM}. We considered as a starting point node 21 - the Theater -, as it is an easily accessible spot by all means of transportation, from public transport (two bus stops are near, with one being right in front of it), to private transport (car parking is available on-site) and walking. As the final destination, we have considered node 22, the Citadel, which is further from other attractions, and less connected, having only one vertice which connects the node associated with it to the graph. 

\begin{figure}[h!]
	\includegraphics[width=0.4\textwidth]{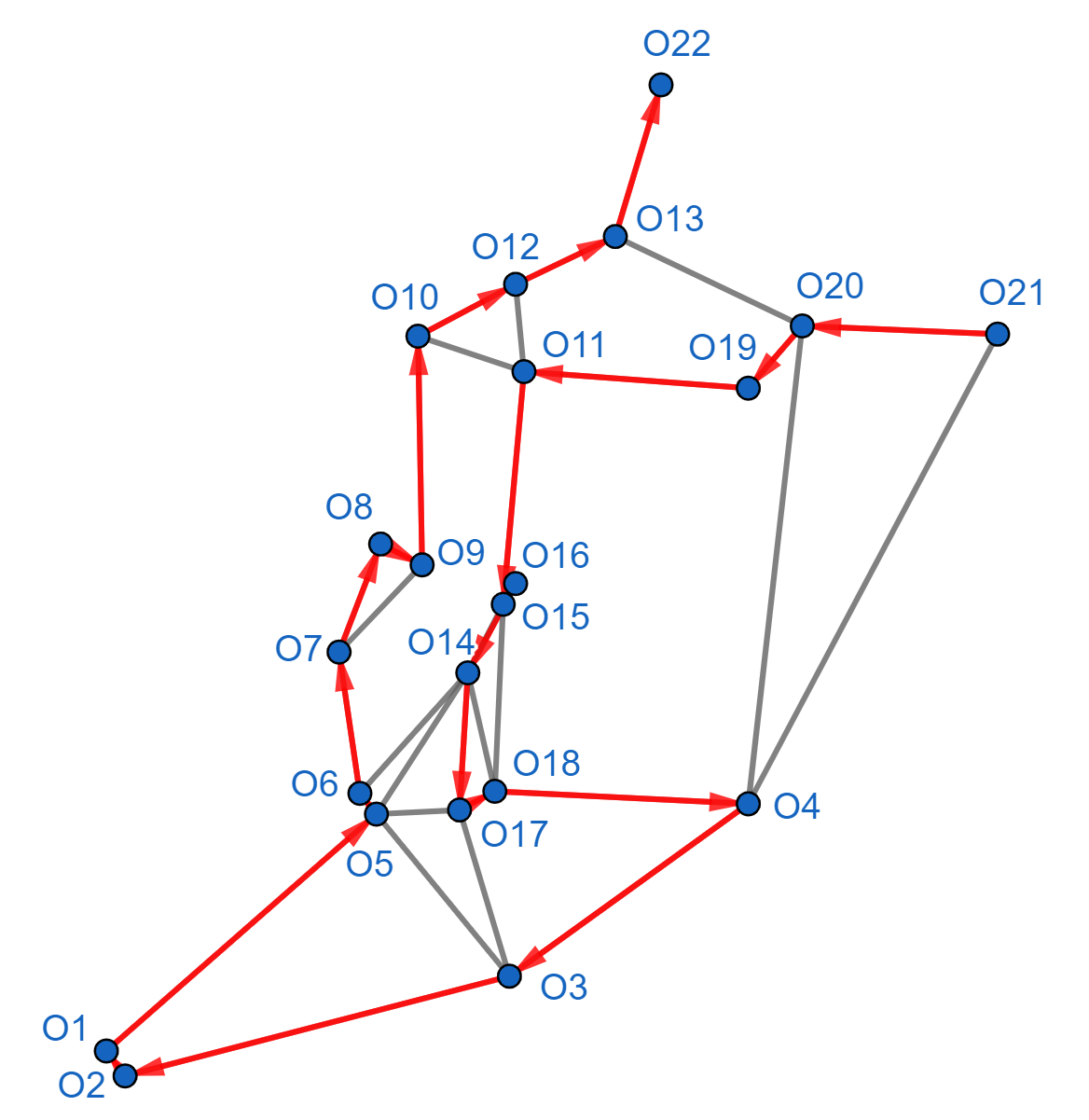}
	\caption{Shortest route to visit all 22 tourist attractions}
	\label{All22}
\end{figure}

Computing the shortest route which passes through all attraction points in the Old City of Brașov we obtain 8 possible itineraries. Measuring the distances in meters, we obtain a significant difference, of over 1 km between the shortest route (7330 m) and the longest (8570 m). The case is similar when calculating the shortest path in minutes (110 minutes), which is 17 minutes less than the longest itinerary. The shortest path, in both cases, starting from the Theater, and ending at the Citadel is as follows:

Theater (21) $\to$ Town Hall (20) $\to$ Art Museum (19) $\to$ Rectorate of Transilvania University of Brasov (11) $\to$ Council Square (15) $\to$ History Museum (16) $\to$ Black Church (14) $\to$ Synagogue (17) $\to$ Rope Street (18) $\to$ Tâmpa Cable Way (4) $\to$ Weaver's Fortress (3) $\to$ Saint Nicholas Church (2) $\to$ First Romanian School (1) $\to$ Șchei's Gate (5) $\to$ Catherine's Gate (6) $\to$ Black Tower (7) $\to$ White Tower (8) $\to$ Graft Fortress (9) $\to$ George Barițiu County Library (10) $\to$ House of Army (12) $\to$ Annunciation Church (13) $\to$ The Citadel (22).

The shortest route which passes through all of the 22 attraction points considered in Brasov is highlighted in red in Figure \ref{All22}.

Although finding the shortest route which passes through all landmarks can be useful, in the context of the Covid-19 pandemic, emphasis must be put on fluidized routes inside a destination to avoid overcrowding places and make it possible for all the social distancing measures imposed by the crisis to be respected.

Thus, we believe that identifying alternative routes, which yield multiple possibilities for travelling inside a destination is much more useful in the Covid-19 context. 

To revive tourism and assure equity from the economic point of view in accordance with the shortening of the tourist stay, we consider that the visitation time for the old centre of Brașov should be of approximately one day. Thus, although the old town of Brașov is not very big, an 8 km route is too much to cover in one day. For this matter, we modified the method to find the route which crosses through a smaller number of attractions.

Taking the target number of attractions to be $10$, which is a reasonable number considering the distances inside Brașov and the mean time spent at one destination, we found all routes which start from node $21$ and end in node $22$. Table \ref{allWith10} shows the result in increasing order of the length of the route in minutes, which represents the cost of following a certain route.

\begin{table}[h!]
	\centering 
	\caption{Possible routes to visit 10 attraction points}
	\label{allWith10}
	\begin{tabular}{| c c c c c c c c c c c | c |}
		\hline
		&   &   &   &   & Route  &   &    &    &    &     & Cost (minutes)\\
		\hline
		21 & 4 &18 &17 & 5 &14 & 15& 11 & 12 & 13 & 22 &   cost=52\\
		\hline
		21 & 4 &18 &17 &14 &15 & 11& 10 & 12 & 13 & 22 &   cost=53\\
		\hline
		21 & 4 &18 &17 &14 &16 & 15& 11 & 12 & 13 & 22 &   cost=53\\
		\hline
		21 & 4 & 3 & 5 & 6 & 14& 15& 11 & 12 & 13 & 22 &   cost=54\\
		\hline
		21 & 4 &18 &14 &16 &15 & 11& 10 & 12 & 13 & 22 &   cost=54\\
		\hline
		21 & 4 & 3 & 5 & 14& 15& 11& 10 & 12 & 13 & 22 &   cost=55\\
		\hline
		21 & 4 & 3 & 5 & 14& 16& 15& 11 & 12 & 13 & 22 &   cost=55\\
		\hline
		21 & 4 & 3 &17 & 5 & 14& 15& 11 & 12 & 13 & 22 &   cost=55\\
		\hline
		21 & 4 & 3 &17 & 18& 14& 15& 11 & 12 & 13 & 22 &   cost=55\\
		\hline
		21 & 4 & 3 &17 & 18& 16& 15& 11 & 12 & 13 & 22 &   cost=55\\
		\hline
		21 & 4 & 3 & 5 & 17& 14& 15& 11 & 12 & 13 & 22 &   cost=56\\
		\hline
		21 & 4 & 3 &17 & 14& 15& 11& 10 & 12 & 13 & 22 &   cost=56\\
		\hline
		21 & 4 & 3 &17 & 14& 16& 15& 11 & 12 & 13 & 22 &   cost=56\\
		\hline
		21 & 4 &18 &16 &14 &15 & 11& 10 & 12 & 13 & 22 &   cost=56\\
		\hline
		21 &20 & 4 &18 &17 &14 & 15& 11 & 12 & 13 & 22 &   cost=57\\
		\hline
		21 & 4 & 3 & 5 & 6 & 7 & 9 & 10 & 12 & 13 & 22 &   cost=58\\
		\hline
		21 &20 & 4 &18 &14 &15 & 11& 10 & 12 & 13 & 22 &   cost=58\\
		\hline
		21 &20 & 4 &18 &14 &16 & 15& 11 & 12 & 13 & 22 &  cost=58\\
		\hline
		21 &20 & 4 &18 &16 &15 & 11& 10 & 12 & 13 & 22 &   cost=58\\
		\hline
		21 &20 & 4 & 3 & 5 &14 & 15& 11 & 12 & 13 & 22 &   cost=59\\
		\hline
		21 & 4 &18 &17 &14 &15 & 11& 19 & 20 & 13 & 22 &   cost=60\\
		\hline
		21 &20 & 4 & 3 &17 &14 & 15& 11 & 12 & 13 & 22 &   cost=60\\
		\hline
		21 &20 & 4 &18 &16 &14 & 15& 11 & 12 & 13 & 22 &   cost=60\\
		\hline
		21 & 4 &18 &14 &16 &15 & 11& 19 & 20 & 13 & 22 &   cost=61\\
		\hline
		21 & 4 & 3 & 5 & 14& 15& 11& 19 & 20 & 13 & 22 &   cost=62\\
		\hline
		21 & 4 &18 &14 & 6 & 7 & 9 & 10 & 12 & 13 & 22 &   cost=62\\
		\hline
		21 & 4 & 3 &17 & 14& 15& 11& 19 & 20 & 13 & 22 &   cost=63\\
		\hline
		21 & 4 &18 &16 &14 &15 & 11& 19 & 20 & 13 & 22 &   cost=63\\
		\hline
	\end{tabular}
\end{table}

\begin{figure}[h!]
	\includegraphics[width=0.4\textwidth]{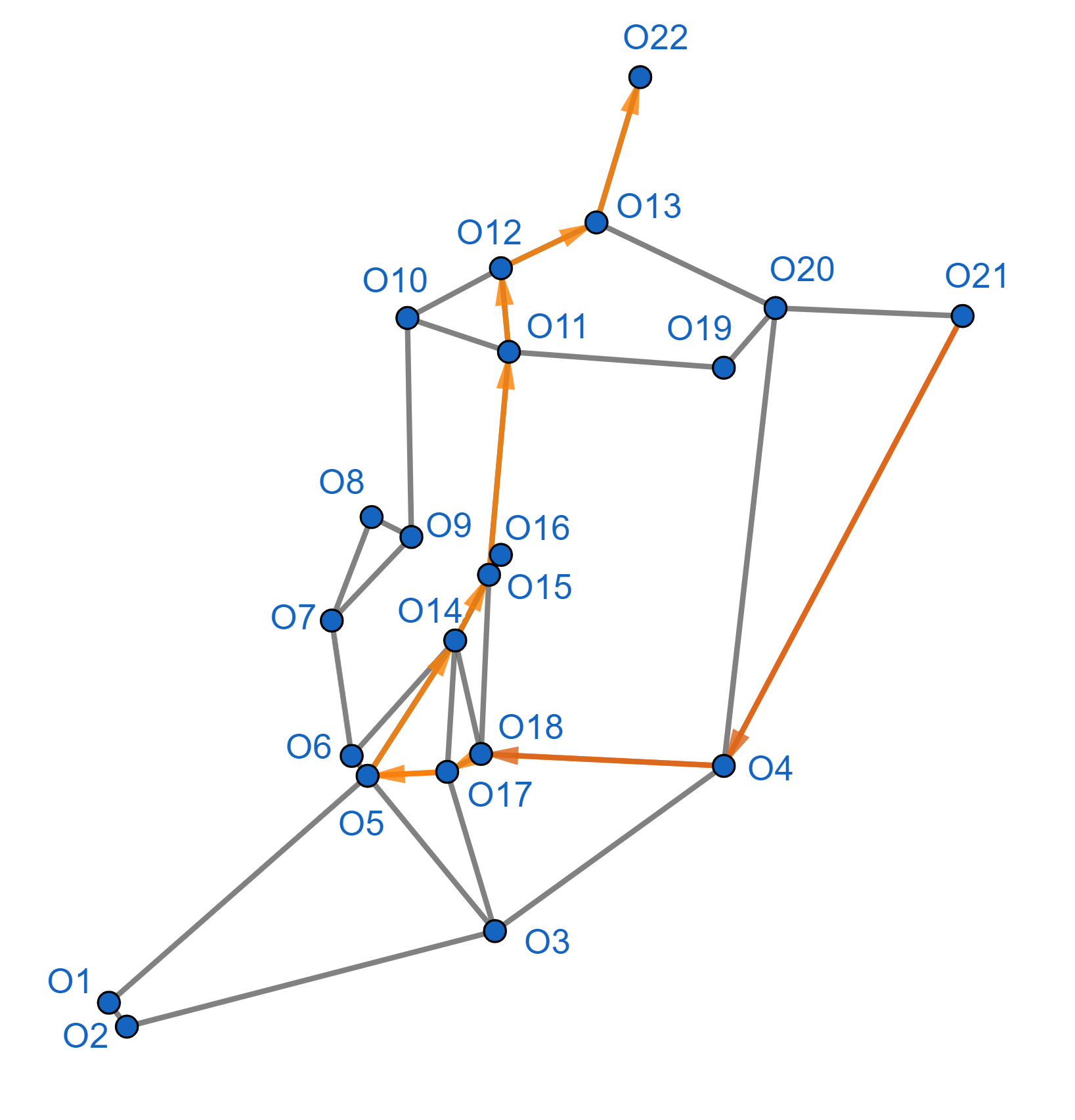}
	\caption{Shortest route to visit 10 tourist attractions}
	\label{Only10}
\end{figure}

The optimal route which passes through 10 attractions is taking 52 minutes, while the longest is a 63-minute walk. The optimal route, for which the cost (the time needed) is minimal is highlighted in orange on the graph in Figure \ref{Only10} and it proposes the following route:

Theater (21) $\to$ Tâmpa Cable Way (4) $\to$  Rope Street (18) $\to$ Synagogue (17) $\to$ Șchei's Gate (5) $\to$ Black Church (14) $\to$ Council Square (15) $\to$ Rectorate of Transilvania University of Brasov (11) $\to$ George Barițiu County Library (10) $\to$ House of Army (12) $\to$ Annunciation Church (13) $\to$ The Citadel (22).

As we can notice from Table \ref{allWith10}, the differences between certain routes are usually not significant, as only a small number of nodes differ. However, if groups with different characteristics, such as age, physical condition, associated diseases, a.o., take different paths, safe tourism can be undertaken as the risk of infection is reduced and tourists' needs are met.

The significance of finding the shortest route (minimal cost - time or distance) is underlined by the existing trend of shortening the length of the tourist stay. The length of the tourist stay is computed using formula
\begin{equation}
\text{LoS} = \dfrac{ONS}{Arr}
\label{len}
\end{equation}
where LoS = Length of Stay, ONS = Over Night Stays, Arr = Arrivals.

\begin{figure}[h!]
	\centering
	\includegraphics[width=8cm]{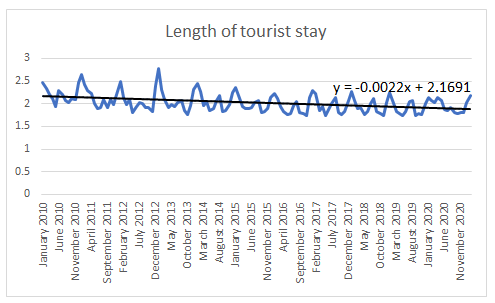}
	\caption{Length of tourist stay in Brasov}
	\label{Length-ef}
\end{figure}

Data for the number of arrivals in Brasov can be accessed from the database TUR104H and the number of overnight stays from the database TUR105H (see \cite{ins}). Based on these data and formula (\ref{len}) we calculated the length of the tourist stay in Brasov. As it can be observed in Figure \ref{Length-ef}, the trend is slightly decreasing. However, the trend line of the real length of visiting the city of Brasov has a much more significant negative slope (i.e. a steeper descending trend) as most tourists chose to find accommodation in Brasov while visiting the surroundings (especially - but not limited - Poiana Brasov, which is a mountain resort situated 10 km from the city).

The importance of identifying not only the fastest route but all routes is underlined by the need to avoid crowded places. Thus, offering multiple itineraries, alternatively available, to groups of tourists is a significant feature.

\subsection{Optimisation of tourist route in the context of Covid-19 restrictions}

In order to adapt the algorithm for the restrictions imposed by Covid-19, we evaluated the crowdedness of the 22 attraction points considered, based on the feature \textit{Popular times}, provided by Google \cite{Google-2}. The data about how busy a certain attraction is corresponds to Winter-Spring 2021, which means that it is calculated during the pandemic when (as can be seen in Table \ref{Covid-ef}) tourism has dropped significantly.

\begin{table}[h!]
	\centering
	\caption{Popular times during Saturdays}
	\label{PopTimes}
	\begin{tabular}{|c|c|c|c|c|c|c|c|c|}
		\hline
		Tourist attraction & 10am & 11am & 12pm & 1pm & 2pm & 3pm & 4pm& 5pm\\
		\hline
		O1: First Romanian School & NB & B & B & VB & VB & B & B & NB \\ 
		\hline
		O2: Saint Nicholas Church$^1$ & NB & B & B & VB & VB & B & B & NB \\  
		\hline
		O3: Weavers' Fortress & NB & NB & NB & B & B & B & B & - \\  
		\hline
		O4: Tâmpa Cable Way & NB & NB & B & B & B & B & B & - \\  
		\hline
		O5: Șchei's Gate & NB & B & B & B & B & B & B & B \\  
		\hline
		O6: Catherine's Gate & NB & NB & B & B & VB & VB & B & B \\  
		\hline
		O7: Black Tower & NB & B & B & VB & VB & VB & B & B \\  
		\hline
		O8: White Tower & NB & NB & NB & NB & B & VB & VB & - \\  
		\hline
		O9: Graft Fortress$^2$ &NB & NB & NB & NB & B & VB & VB & - \\  
		\hline
		O10: George Barițiu Library$^3$ & B & B & VB & VB & VB & VB & VB & VB \\  
		\hline
		O11: Rectorate$^3$  & B & B & VB & VB & VB & VB & VB & VB \\  
		\hline
		O12: House of Army$^4$ & NB & NB & B & B & B & B & B & B\\  
		\hline
		O13: Annunciation Church & NB & NB & NB & NB & NB & NB & NB & NB \\  
		\hline
		O14: Black Church & NB & NB & B & B & B & B & B & NB \\  
		\hline
		O15: Council Square  & NB & NB & B & B & B & B & B & B \\ 
		\hline
		O16: History Museum$^5$ & NB & B & VB & VB & VB & VB & B & NB\\  
		\hline
		O17: Synagogue$^6$ & NB & B & B & B & B & B & B & B \\  
		\hline
		O18: Rope Street & NB & B & B & B & B & B & B & B \\  
		\hline
		O19: Art Museum & NB & NB & B & B & B & NB & NB & NB \\  
		\hline
		O20: Town Hall & NB & B & VB & VB & VB & VB & B & NB \\  
		\hline
		O21: Theater* & - & - & - & - & - & - & - & - \\  
		\hline
		O22: The Citadel & NB & B & VB & VB & VB & B & B & - \\  
		\hline
	\end{tabular}
\flushleft{\small{ NB - \textit{not busy}; B - \textit{busy}; VB - \textit{very busy}\\ $^1$ - approximated from \textit{First Romanian School} \\ $^2$ - approximated from \textit{White Tower}\\ $^3$ - approximated from \textit{Livada Postei}\\ $^4$ - approximated from \textit{Nicolae Titulescu Park} \\$^5$ - approximated from  \textit{Town Hall}\\ $^6$ - approximated from \textit{Rope Street}}}
\end{table}

In order to avoid overcrowding and to ensure that social distancing measures are respected, we impose certain additional restrictions to the algorithm:
\begin{itemize}
	\item[-] if at a certain time an attraction is very busy (VB), we block that objective for that specific time;
	\item[-] if an attraction is not busy (NB), that node is preferred for the route and included in the final route; 
	\item[-] at least one preferred node is included in the final route.
\end{itemize}

While the first restriction ensures that the attractions are not overcrowded, the second and third restrictions ensure that some less-visited points are included in the tourist route, thus providing equity among attractions and a revival of secondary attractions.

Taking into account the data from the Table \ref{PopTimes}, we determined the best route for specific times. Since most attractions are not too busy at 10 am, 11 am and 5 pm, we decided to build only the optimal routes for the remaining time intervals. The computed results for the optimal route for each time frame are presented in Table \ref{routes1}. The cost of each route is calculated in meters.

\begin{table}[h!]
	\caption{Optimal routes at different hours for Saturdays in Brasov}
	\label{routes1}
	\begin{tabular}{ccc}
		\textbf{Time} & \textbf{Route} & \textbf{Cost}\\
		\hline
		12 am & Theater - Tampa Cable Way - Weaver's Fortress - & \\    & Synagogue - Rope Street - Council Square - & 2990 \\
		&House of Army - Annunciation Church - The Citadel & \\
		13 am & Theater - Tampa Cable Way - Weaver's Fortress - & \\    & Schei's Gate - Catherina's Gate - Graft Fortress - & 2950 \\
		&House of Army - Annunciation Church - The Citadel & \\
		14 am & Theater - Tampa Cable Way - Weaver's Fortress - & \\    & Schei's Gate - Graft Fortress - House of Army  & 2700\\
		& Art Museum - Annunciation Church - The Citadel & \\
		15 am & Theater - Tampa Cable Way - Weaver's Fortress - & \\ & Schei's Gate - House of Army - Art Museum - & 2700 \\ & Annunciation Church - The Citadel & \\
		&House of Army - Annunciation Church - The Citadel & \\
		16 am & Theater -  Town Hall - Art Museum - &\\ & House of Army - Annunciation Church - The Citadel  & 2880\\
	\end{tabular}
\end{table}

In order to test the model for optimizing the tourist route presented in this paper, we used a qualitative research based on a sample of 15 volunteers who were willing to get involved in testing the model.  

The objectives of the qualitative research were:

\begin{itemize}
	\item[$Q_1$] identifying the level of congestion on the five routes proposed in the specified five 5 time frames;
	\item[$Q_2$] discovering some less popular attractions in the five routes, or located in areas adjacent to them;
	\item[$Q_3$] determining the degree of satisfaction of the volunteers regarding the tourist experience related to the proposed route.
	\item[$Q_4$] assessing the utility of implementing the method in the future.
\end{itemize}

The qualitative research was conducted in Brasov in April 2021. The volunteers were both residents of Brasov (7 volunteers) and tourists visiting Brasov (8 volunteers). Each volunteer filled a questionnaire after completing the route proposed based on the associated time frame. The tourists were recruited in the vicinity area of crowded attractions in Brasov. 

The questionnaire was composed of 12 questions, included in a previously created interview guide (see Appendix). We present the most important conclusions that were drawn from the interpretation of the volunteers' answers. 

The distribution on the routes was as follows:
\begin{itemize}
	\item[-] 4 participants followed Route 1 and started the tour at 12 pm;
	\item[-] 2 participants followed Route 2 and started the tour at 1 pm;
	\item[-] 3 participants followed Route 3 and started the tour at 2 pm;
	\item[-] 3 participants followed Route 4 and started the tour at 3 pm;
	\item[-] 3 participants followed Route 5 and started the tour at 4 pm;
\end{itemize}

Most participants answered that they did not encounter overcrowded places along their route (66.7\%). All participants who chose Route 5 (3 participants) answered that they encountered overcrowded places and pointed to the Town Hall as one of the busiest attractions. However, almost all participants (14 out of 15) described the suggested route as helpful (5) and very helpful (9) in maintaining social distancing measures imposed by the authorities. Thus, the method proposed can be a useful tool in tourist route planning in the context of the Covid-19 pandemic (see objective $Q_1$).

Among the people interviewed, 60\% (9) answered that they have discovered new, less-visited places. Among the answers, the \textit{Annunciation Church} (6 answeres), \textit{House of Army} (4 answeres)  and \textit{Art Museum} (4 answeres) have been discovered by most participants. Concerning attractions that were not listed on the routes, but which were adjacent to the routes, tourists discovered \textit{the caves and galleries in the Graft area}, \textit{the trees and shrubs and the wood carvings in the park near Schei s Gate}, \textit{the ice ring near the Tampa cable way} and \textit{the fresh spring scenery of the forest at the bottom of Tampa}. From the answers received, we can conclude that the implementation of this method brings a revival of less visited tourist attractions in a destination, thus ensuring a fair distribution of tourists inside an urban destination (see objective $Q_2$). 

In terms of satisfaction, 13 participants described the route as satisfactory (6) and very satisfactory, while 2 participants found the route to be neutral (1) and a little satisfactory. Also, 13 participants admitted that following the specified routes optimized their visit to a great extent in terms of its duration, and the same number of respondents agreed that the route has been very efficient (6) and efficient (7) (see objective $Q_3$).

At the final question, which evaluates the utility of implementing the method in the future, 66.7\% (10) of volunteers said that they would definitely use a tool to suggest the optimal route, while the rest of 26.7\% (4) answered \textit{Maybe} (see objective $Q_4$).

We believe that the results obtained from this first pilot study on the method proposed are satisfactory and prove that the method would be beneficial for tourists as well as for the community. The routes proposed proved to be less overcrowded and brought to the spotlight less-visited sights inside Brasov, a tourist city that was greatly affected by the pandemic.

\section{Discussions}

Optimizing the tourist routes within a destination in the context of COVID-19 is one of the main challenges which should be resolved by a joined effort from authorities and tourism industry.

In the context of the COVID-19 crisis, the length of the tourist stay is expected to be significantly reduced and thus, this already existing trend will be accelerated. Another main effect of the present pandemic, which dictates the imperative need to create specific routes inside a destination, is that the revival of tourism should start locally to reduce the risk of spreading the virus, while positively contributing to the local economy.

On the other hand, the COVID-19 crisis has brought to the spotlight the significant issue of overcrowded places, which pose an imminent threat as they are potentially outbreak points. Our proposal of an optimized route in the tourist destination of Brașov aims to redirect people on slightly different routes to partially solve this problem.

The method can be efficiently used by authorities to create an online application with interfaces both for web and for phone. Such implementation of the method proposed can be an efficient tool in creating a safe environment for tourism, which will bring benefits to all levels of tourism. Another benefit of implementing our method in an application available for tourists is that tourists with less available time for visiting can choose the number of attraction points they want to visit and a starting and ending point and discover a suitable route that would fit their time frame.

We believe that in addition to the many problems brought by Covid-19, the global pandemic has highlighted a positive side, which demands a reinvention of all tourist destinations due to an imperative adaptation to the restrictions imposed by the pandemic. This could be reached with the help of creativity and with support from all stakeholders which ought to start drawing a new type of tourism in accordance with the obvious needs of tourists and investors, but also with the protection needs of the natural tourist potential of the destination - that means a true responsible tourism.

The method should be incorporated in a location-based application that should get real-time data about the number of people present at a certain location at a certain point. Based on this data, the user will be able to choose the optimal length of the visit, and (or), the number of attractions that should be visited, as well as some additional information (existing medical conditions, age, or other limitations) which will allow the application to choose the best route.

According to Su et al. (see \cite{Su}), in order to innovate the services included in the tourist offer and to ensure an enhanced attractivity of the touristic products offered inside a certain destination, the tourist offer must be correlated with the image of the destination, as it results from its visitors’ descriptions of the experiences welcomed there. We find the idea of creating \emph{an official website where tourists can share an edited view of their personal life with people they select so that tourists are more likely to transform utilitarian well-being into intrinsic motivation and thus commit to their activities} (see \cite{Su}) particularly interesting as the customization of tourist routes inside a destination that we propose in this paper should be realized based on certain particularities definitory to a specific group interested to visit the attraction points discussed. We believe that the tourist routes adapted to the characteristics of tourists who are interested in visiting the old town of Brașov, that we propose in this paper, would encourage the active involvement of the tourists in determining the most appropriate routes which comply with certain requirements previously stated by tourists who have visited the old town of Brașov (safety, time spent, intrinsic motivation, a.o.).

\section{Conclusions}

The implementation of the method inside tourist destinations can bring an important advantage in transforming a destination into a safer one in times of Covid-19 and post-Covid-19. Moreover, the existing trend of shortening the length of tourist stay underlines the necessity of better planning the visit of a certain destination, from the tourist's point of view, as well as an optimal organization of the tourist routes inside a destination in order to offer equity for all beneficiaries of the tourist industry. 

We consider that the proposed method of optimizing the tourist routes in the context of the pandemic will be a tool with practical applicability for implementing a sustainable development strategy for tourism industry in Brasov. This could be really useful for DMO (Destination Management Organisation)-Brasov, in the process of developing and implementing a proper strategy, which should take into account the particular interests of all stakeholders (tourists interested in both the tourist experience and the safety of their families, local authorities, investors, local community, environment, different NGO's).

The implementation of the present method aims to cover all three instances of sustainable tourism. From the economic point of view, well-organized tourist routes will bring more income to the community while assuring a fair chance for development to a greater amount of small businesses. Moreover, planning the visiting route for one day will positively affect the time budget, increasing the possibility to visit more destinations from Brașov county, thus improving the regional tourism. From the social point of view, reducing agglomeration in certain spots by redirecting some tourists on different routes will result in an increase of small business which provides jobs for people. Personalized routes are also advantageous as they give the possibility to multiple categories of people to visit the objectives. From this point of view, in the context of the current pandemic, people with existing health conditions, which represent a risk category, can pick a route based on attractions that are of interest to them, located exclusively in the open. Finally, from the ecologic point of view, creating different routes for visitors can reduce the agglomeration on some spots which can wear out if they are overcrowded.

\vspace{6pt}

\newpage
\appendix 
\textbf{Appendix}

\begin{enumerate}
	\item[1.] Do you live in Brasov?
	\begin{itemize}
		\item[$\circ$] Yes
		\item[$\circ$] No
	\end{itemize}
	\item[2.] What is your age?
	
	$\dots\dots\dots\dots$
	
	\item[3.] What was the time frame of your visit in Brasov?
	\begin{itemize}
		\item[$\circ$] 12pm
		\item[$\circ$] 1pm
		\item[$\circ$] 2pm
		\item[$\circ$] 3pm
		\item[$\circ$] 4pm
	\end{itemize}
	\item[4.] What route did you follow?
	\begin{itemize}
		\item[$\circ$] Route 1: Theater - Tampa Cable Way - Weaver's Fortress - Synagogue - Rope Street - Council Square - House of Army - Annunciation Church - The Citadel
		\item[$\circ$] Route 2: Theater - Tampa Cable Way - Weaver's Fortress - Schei's Gate - Catherine's Gate - Graft - House of Army - Annunciation Church - The Citadel
		\item[$\circ$] Route 3: Theater - Tampa Cable Way - Weaver's Fortress - Schei's Gate - Graft Fortress - House of Army - Art Museum - Annunciation Church - The Citadel
		\item[$\circ$] Route 4: Theater - Tampa Cable Way - Weaver's Fortress - Schei's Gate - House of Army - Art Museum - Annunciation Church - The Citadel
		\item[$\circ$] Route 5: Theater -  Town Hall - Art Museum - House of Army - Annunciation Church - The Citadel
	\end{itemize}
	\item[5.] Have you discovered new, less-visited tourist attractions?
	\begin{itemize}
		\item[$\circ$] Yes
		\item[$\circ$] No
		\item[$\circ$] Maybe
	\end{itemize}
	\item[6.] If the answer to the previous question was yes, please indicate those newly  revealed objectives?
	
	$\dots\dots\dots\dots\dots\dots\dots\dots\dots\dots\dots\dots\dots\dots\dots\dots$
	\item[7.] Have you encountered overcrowded places along your route?	
	\begin{itemize}
		\item[$\circ$] Yes
		\item[$\circ$] No
	\end{itemize}
	\item[8.] How satisfied are you with the route you followed? (1- not al all, 5 - very satisfied)
	\begin{itemize}
		\item[$\circ$] 1
		\item[$\circ$] 2
		\item[$\circ$] 3
		\item[$\circ$] 4
		\item[$\circ$] 5
	\end{itemize}
	\item[9.] How much do you think following the route optimized your visit in Brasov in terms of its duration? (1- not al all, 5 - very much)
	\begin{itemize}
		\item[$\circ$] 1
		\item[$\circ$] 2
		\item[$\circ$] 3
		\item[$\circ$] 4
		\item[$\circ$] 5
	\end{itemize}
	\item[10.] How much do you think the route chosen has helped you maintain social distancing? (1- not al all, 5 - very helpful)
	\begin{itemize}
		\item[$\circ$] 1
		\item[$\circ$] 2
		\item[$\circ$] 3
		\item[$\circ$] 4
		\item[$\circ$] 5
	\end{itemize}
	
	\item[11.] How effective do you think the route has been? (1- not al all, 5 - very much)
	\begin{itemize}
		\item[$\circ$] 1
		\item[$\circ$] 2
		\item[$\circ$] 3
		\item[$\circ$] 4
		\item[$\circ$] 5
	\end{itemize}
	
	\item[12.] Would you use a tool suggesting an optimal route in your future visiting tours?
	
	\begin{itemize}
		\item[$\circ$] Yes
		\item[$\circ$] No
		\item[$\circ$] Maybe
	\end{itemize}
\end{enumerate}

%


\end{document}